\documentclass[runningheads]{llncs}
\usepackage[T1]{fontenc}
\usepackage{newtxtext}       %
\usepackage[varvw]{newtxmath}       

\usepackage{graphicx}

\usepackage[utf8]{inputenc}
\usepackage[T1]{fontenc}
\usepackage{booktabs}
\usepackage{array}
\usepackage{makecell}
\usepackage{siunitx}
\usepackage{comment}
\usepackage[table]{xcolor}
\begin{document}

\title{Anonymous in Name Only: A Critical Review and a Reidentification Exercise on the European Commission’s 
Anonymization Measures for Google Search Data under Article 6(11) of the DMA}

\titlerunning{A Critical Review on the EC’s Anonymization Measures for Google Search Data}
%
\author{David Sánchez\inst{1} \and
Benet Manzanares-Salor\inst{1} \and
Josep Domingo-Ferrer\inst{1,2}}
\authorrunning{D. Sánchez {\em et al.}}
%
\institute{Department of Computer Science and Mathematics, ComSCIAM-Center for Computational Science and Applied Mathematics, CYBERCAT-Center for Cybersecurity Research of Catalonia, Universitat Rovira i Virgili, Av. Pa\"{\i}sos Catalans, 26, Tarragona, 43007, Catalonia. \\
  \email{\{david.sanchez, benet.manzanares, josep.domingo\}@urv.cat}
  \and
  \'Ecole normale sup\'erieure, Universit\'e PSL, Paris, France}

\maketitle              
\begin{abstract}
In July 2026, the European Commission (EC) set out the anonymization measures Alphabet must implement to share Google Search data with competing search engines under Article 6(11) of the Digital Markets Act. As the first concrete illustration of how the EC operationalizes anonymization under the EU digital rulebook, this 
methodology could become a regulatory benchmark. In this paper, {\em we show why it should not}. We critically review the anonymization measures proposed by the EC, from the characterization of residual risk to the suppression thresholds and the reliance on contractual safeguards. For each of those issues, we highlight significant conceptual flaws, and, where applicable, we illustrate these flaws with empirical observations from real Google Search data provided by Alphabet and protected under the EC’s measures. Importantly, we present examples of low-effort reidentification, record linkage, and confidential attribute inference achieved through straightforward and automatable attack patterns. Our findings show that the approach of the EC relies on unjustified heuristics rather than formal privacy guarantees, underestimates the risks inherent in free-text search queries, and sacrifices data utility without achieving meaningful protection. We also demonstrate that the proposed measures do not meet any of the three anonymization criteria established by the European Data Protection Board (EDPB): no record isolation, no linkage, and no inference. Practitioners and regulators 
are therefore cautioned against treating these measures as a sound anonymization practice.

\end{abstract}
\section{Introduction}

On 16 July 2026, the European Commission (EC) released its findings in the case DMA.100209 – SP against Alphabet \cite{EC}, which sets out the measures that Alphabet will be required to implement to comply with its data sharing obligations under Article 6(11) of the Digital Markets Act (DMA) \cite{DMA}. In particular, it requests Alphabet to share Google Search data with other search engines after being subjected to \emph{anonymization}.  

This case is particularly significant because it provides the first concrete illustration of how the EC operationalizes data anonymization under the EU digital rulebook \cite{rulebook}, whose fundamental pillars include the General Data Protection Regulation (GDPR) \cite{GDPR}. More broadly, the proposed anonymization measures arguably represent the most comprehensive attempt by any regulator worldwide to establish a framework for privacy-preserving data sharing of online search data. As such, this case sets an important precedent and is likely to have a substantial influence on the future adoption and implementation of anonymization practices by both regulators and practitioners.

The EC defines both technical and contractual measures for anonymization. In this paper, we argue that these measures fail to render the data anonymous and do not satisfy any of the criteria for effective anonymization set out by the European Data Protection Board (EDPB). Specifically, we show that the EC's approach is unrealistically optimistic, relies on heuristics rather than providing guarantees, and lacks technical justification. We illustrate these flaws with empirical findings obtained from exhaustive six-month samples of Google Search data provided by Alphabet and protected under the EC's measures. \emph{These are precisely the same data that competing search engines would have access to.}
Importantly, we report numerous examples of low-effort reidentification, record linkage, and confidential attribute inference achieved through straightforward and automatable attack patterns.

The remainder of the paper is organized as follows. Section~\ref{sec:legal} summarizes the applicable legal context. Section~\ref{sec:anonymization} reviews the fundamental principles of data anonymization. Section~\ref{sec:measures} details the anonymization measures proposed by the EC. Section~\ref{sec:review} critically examines these measures. Section~\ref{sec:review} details attack patterns and examples of reidentification and disclosure. Section~\ref{sec:guidelines} explains why EC measures do not meet the anonymization criteria set out by the EDPB. Finally, the last section presents the conclusions of our analysis.

\section{Legal context}
\label{sec:legal}

Article 6(11) of the DMA \cite{DMA} states that: ``The gatekeeper shall provide to any third-party undertaking providing online search engines, at its request, with access on fair, reasonable, and non-discriminatory terms to ranking, query, click and view data in relation to free and paid search generated by end users on its online search engines. Any such query, click and view data that constitutes personal data shall be anonymised.''

The goal, as expressed in Recital (61) of the DMA, is to ``ensure the protection of the personal data of end users, including against possible reidentification risks, by appropriate means, such as anonymization of such personal data, without substantially degrading the quality or usefulness of the data. The relevant data is anonymized if personal data is irreversibly altered in such a way that information does not relate to an identified or identifiable natural person or where personal data is rendered anonymous in such a manner that the data subject is not or is no longer identifiable.''

Because the DMA and the GDPR must be read together (the concept of personal data comes from the GDPR, see Article 2(25) of the DMA), the anonymization standard should be coherent with the definitions of the GDPR and Recital (26) \cite{GDPR}. 

Specifically, the GDPR defines personal data as any information relating to an identified or identifiable living natural person. This includes both facts and opinions about a person. The GDPR (Art. 9) also makes a special point on ``sensitive personal data'', which accounts for personal data revealing racial or ethnic origin, political opinions, religious or philosophical beliefs,  trade union membership, and data concerning health, sex life or sexual orientation, whose processing is strictly prohibited. 

Finally, Recital (26) of the GDPR states that ``to determine whether a natural person is identifiable, account should be taken of all the means reasonably likely to be used, such as singling out, either by the controller or by another person to identify the natural person directly or indirectly. To ascertain whether means are reasonably likely to be used to identify the natural person, account should be taken of all objective factors, such as the costs of and the amount of time required for identification, taking into consideration the available technology at the time of the processing and technological developments.'' 

Very recently, the EDPB has released Guidelines 02/2026 on Anonymisation \cite{edpb}, which define three criteria that should be met for personal data to be regarded as anonymous:
\begin{itemize}
\item \emph{No Record Isolation}. This criterion is met ``if the data does not contain a unique combination of attribute values that relate to a single individual.'' If the data violate this criterion, at least it should be ensured that the isolated records do not allow for singling out, that is, that the individuals are distinguished from others by matching those records with the individual's corresponding attributes.  
\item \emph{No Linkage}. This criterion is met ``if the data does not contain an individual's record which could be linked to another record which also relates (with certainty or high likelihood) to that same individual.'' 
\item \emph{No Inference}. This criterion is met ``if no specific and meaningful inference can be drawn from the given data. Inferences may be based on the given data and/or on additional information, including information about the anonymization process.''
\end{itemize}

To align the GDPR (which restricts transfers of personal data) and the DMA (which requires gatekeepers to transfer data to facilitate competition), the EC and the European Data Protection Board have published joint draft guidelines that describe the interplay between both regulations~\cite{interplay}. These draft guidelines were released for public consultation on 9 October 2025, and the consultation closed on 4 December 2025. However, no final version of these guidelines has yet been adopted.

\section{Background on data anonymization}
\label{sec:anonymization}

Privacy protection began in the 1970s in official statistics as a heuristic and empirical approach: a statistical disclosure control (SDC) method \cite{sdc-book}, such as data suppression, sampling, or noise addition, is tried and evaluated iteratively with varying parameter values until a parameter choice is empirically found that preserves sufficient analytical utility while bringing the disclosure risk below a certain threshold. 

However, the rapid growth of the Web and the widespread collection and exchange of personal data that began in the late 1990s exposed the limitations of heuristic SDC methods \cite{Sweeney2000}. This gave rise to a fundamentally different paradigm based on \emph{privacy models}. Instead of empirically evaluating disclosure risk after protection, privacy models specify an \emph{ex ante} privacy condition that the released data must satisfy. This condition is expressed and verified through precise mathematical definitions, rather than informal empirical evidence, and therefore does not require an \emph{ex post} risk assessment.

The first privacy model was $k$-anonymity~\cite{k-anonymity}, which was specifically designed to anonymize \emph{microdata sets} (\emph{i.e.}, datasets consisting of records containing attribute values relating to individuals) for the release or sharing of data. This is arguably the most common and widely sought use case of data anonymization~\cite{del2026differential}.

Under $k$-anonymity, privacy-sensitive attributes are classified as follows:
\begin{itemize}
\item {\em Identifiers} are attributes that allow unambiguous reidentification of the individual to whom they refer. Examples include social security numbers or passport numbers. 
Identifiers should always be suppressed in anonymized data releases.
\item {\em Quasi-identifiers} (QI) are attributes that do not lead
to reidentification when taken one by one, but a combination of them may if: i) the combination of values on a target individual is known by the intruder, and ii) this combination is unique in the population.
For example, \cite{Sweeney2000} shows that 87\% of the population in the United States can be unambiguously identified once their 5-digit ZIP code, birth date, and sex are known. QIs should be either suppressed or masked (\emph{e.g.}, generalized, swapped, perturbed, etc.). The latter is preferable, as it provides better preservation of data utility.
\item{\em Confidential attributes} hold confidential
information on the individual that shall not be disclosed, such as 'sensitive personal information' types defined in the GDPR. 
\end{itemize}

The attribute types described above give rise to two main types of disclosure~\cite{sdc-book}:
\begin{itemize}
\item {\em Identity disclosure}, also known as {\em reidentification}. The attacker links an otherwise unidentified record to the identity of the individual to whom it refers by exploiting (quasi-)identifying attributes and background knowledge on the individual.
\item {\em Attribute disclosure}. The attacker infers with high certainty the value of a confidential attribute for a target individual from the released data.
\end{itemize}
Both types of disclosure can occur, one of them, or none of them.
The \emph{no record isolation} and \emph{no linkage} principles in the EDPB guidelines~\cite{edpb} are meant to protect against reidentification, whereas the \emph{no inference} principle is meant to protect against attribute disclosure. 

$k$-Anonymity assumes that the identifiers are suppressed
from the released data, but the values of the quasi-identifier attributes can be known by an attacker,
who can use them for reidentification. Consequently, $k$-anonymized microdata must be computed so that each combination of values of the quasi-identifier
attributes is shared by $k$ or more records. 
In this way, for any $k>1$, \emph{unequivocal} reidentification of individuals from the $k$-anonymous data is no longer possible --as no unique records remain in the dataset--, while the \emph{probability} of successful but uncertain reidentification is bounded by $1/k$. 

$k$-Anonymity is not tied to a specific masking mechanism, and can be achieved by masking quasi-identifiers through a variety of masking methods, including non-perturbative (\emph{i.e.}, truthful) generalization and suppression --the original approach in \cite{samarati1998protecting,k-anonymity}-- or perturbative aggregation --the microaggregation approach in \cite{domingo2005ordinal}.

Following the AOL search data reidentification scandal \cite{aol}, extensive research on the anonymization of search queries has been conducted over the past two decades 
\cite{terrovitis08,He09,Navarro2012,Batet2013,Sanchez18}. A fundamental assumption, consistently shared across this body of research, is that not only search-query metadata (\emph{e.g.}, IP addresses, timestamps, device types, etc.) can be (quasi-)identifying, but also the search-query text itself. Indeed, users' queries may contain pieces of information that can directly, or in combination, reveal their identity, while also directly or indirectly disclosing ``sensitive'' information types covered by the GDPR. In this context, applying $k$-anonymity on the entire content of each search record (metadata + query text) constitutes a natural solution, as it \emph{guarantees} that no search record remains unique in the anonymized dataset and, consequently, that unequivocal reidentification and 
attribute disclosure based on the released data are no longer possible.
 
Alphabet's proposal for Google search data anonymization consisted in enforcing $k$-anonymity with $k=30$ on the queries and with $k=5$ on the entire search record (metadata + query). In the first version of their proposal, search queries that do not satisfy the thresholds are suppressed. In a second version, these queries 
are replaced by semantically similar queries that already 
satisfy the thresholds. Both versions were rejected by the EC, which argued that ``Alphabet has not considered all appropriate means to anonymize Google Search Data, and that alternative technical means that have a less detrimental impact on data quality and usefulness are available'' (Recital (583)). 
The EC also criticizes that Alphabet's proposal masks ``all personal data from query text -- even if the personal data is not associated with the end user that issued the query'' (Recital (574)).

\section{Overview of the EC measures on anonymization}
\label{sec:measures}

Alphabet is asked to share Google Search data daily at the record level, with each record containing a user's query and its associated metadata. The search data should be protected by the measures proposed by the EC, whose goal is to ``ensure anonymization of end users' personal data'', and which consist of technical and contractual measures (Recital (593)).

Under the EC's technical measures, several record attributes are removed: Google user identifiers, precise timestamps (although the day of the query is retained), result screen layout information, and queried images if any, which are replaced by a placeholder while any associated query text is retained. What remains is the query text itself, together with metadata fields specifying the user's location (through a NUTS3 code), the inferred language of the query, the user's device type, and the aforementioned timestamp at day level, as well as the URLs of all standard search results.

On these remaining data, the EC measures apply \emph{query suppression} and \emph{metadata generalization}. Query suppression operates through two thresholds. First, an \emph{entity-based threshold} checks whether every query entity (direct identifiers such as names, addresses, phone numbers, etc., or individual words that do not fall into the previous categories) is found in a pre-built allow list of entities appearing in queries submitted by at least 50 signed-in users over the past 13 months. The allow list is updated monthly, and allowed entities remain on the list for five years. Second, a \emph{length-based threshold} checks whether the query text is shorter than 95\% of distinct queries issued in 2025 in the query's inferred language\footnote{The calculation of this threshold was one of the few changes the EC implemented following the public consultation on the preliminary measures, held in April 2026 (\url{https://digital-markets-act.ec.europa.eu/dma100209-consultation-proposed-measures-google-search-data-sharing\_en}). In the preliminary version, the 95\% length-based threshold was to be updated weekly using data from the preceding week.}. Any search record whose query does not meet either threshold 
is suppressed\footnote{Search records contain two query versions: the \emph{initial query} submitted by the user and the \emph{modified query} actually executed after applying suggested query changes, if any. Each version is assessed independently against the thresholds, and if at least one version meets the thresholds, the search record is retained but only with the version(s) that pass the thresholds. The entire search record is suppressed only if neither version meets the thresholds.}.

For search records that pass the suppression thresholds, metadata generalization ensures that each combination of inferred query language, location, and device type is shared by at least 1,000 signed-in users. The only metadata attribute subject to generalization is location. If fewer than the required number of user queries share the same combination of metadata with the location generalized to the country level, the search record is suppressed. The query timestamp is not considered in the metadata combination and is retained at day-level granularity.

These technical measures are complemented by contractual provisions that essentially prohibit data recipients from attempting reidentification. Their necessity is justified by the requirement to ``mitigate the residual likelihood of identification that remains after applying the technical measures'' (Recital (789)).

\section{Critical review}
\label{sec:review}

First, it is important to highlight some characteristics of the Google Search dataset that have a significant bearing on its privacy risks. 
The EEA has a population of 458 million and  
Google Search has a dominant position in the EEA, with a market share of approximately 
89.69\%, while Bing's share is 4.74\%, Yandex's 2.26\%, Yahoo!'s 1.37\%, and DuckDuckGo's 0.84\% \cite{marketshare}. 
At the same time,  
according to Eurostat~\cite{eurostat}, 95\% of people aged 16-74
years use the Internet and web search is basic to them.
Consequently, the Google Search data to be shared by Alphabet constitute a nearly exhaustive sample of the EEA population.

This has important implications from a privacy perspective. With a nearly exhaustive sample of the population, successful reidentification based on 
quasi-identifiers\footnote{If reidentification takes place using actual identifiers that unduly appear in the anonymized data, then it is unequivocal even if the anonymized data are a small sample of the population.} or accurate attribute inference constitutes unequivocal disclosure of information about the individuals concerned in most cases. This is because the attacker can be almost sure that the target individual is represented in the dataset and, if the matched record is unique, the resulting disclosure is unambiguous. This makes the Google Search dataset particularly privacy-sensitive and raises the bar for any anonymization mechanism intended to protect its users.

On the other hand, the scenario of data sharing, even with controlled recipients, differs fundamentally from conventional static and monolithic data releases, such as those offered by statistical offices \cite{muralidhar2025protecting}. In the current case, search data are dynamically generated as a continuous stream and at a volume so vast that they effectively escape any kind of human oversight. Consequently, any anonymization measures must be self-sufficient, as there will be no opportunity for meaningful \emph{ex post} assessment before the data are shared. Moreover, such measures should be robust not only against classic privacy threats, but also against emerging attacks and analytical capabilities enabled by the rapid development of AI. 

Finally, Google Search data accumulate over time (years), giving potential attackers an increasingly long window in which to aggregate users' search records.

In this section, we argue that the data protection needs for this kind of data sharing are not met by the heuristic anonymization approach proposed by the EC, which is unrealistically optimistic and lacks technical justification. In particular, having discarded the \emph{ex ante} formal privacy guarantees offered by privacy models such as $k$-anonymity, the EC has instead chosen to base its anonymization proposal on assumptions and empirical observations that, as discussed below, are at best questionable and, in many cases, incorrect.
 
Where applicable, we illustrate the flaws in the EC's proposal with empirical observations of real-world data provided by Alphabet, comprising six months of exhaustive Google Search data from several EEA countries to which the EC's technical measures had been applied (that is, the same data that competing search engines would have access to). We also report detailed examples of reidentification and disclosure in Section \ref{sec:examples}. Our analyses focus on data from Ireland (comprising 4.87 million search records, mostly in English) and from Spain (comprising 34.62 million search records, with Spanish as the dominant language). We also validated our findings against the corresponding allow list.

\subsection{Inappropriate assessment of privacy risks}
\label{sec:risk}

In Recital (619) the EC states that it ``assesses risks in light of the framework proposed in the Expert Report on Anonymisation according to which risks are considered beyond residual level after application of technical measures if they meet any of the following criteria:
\begin{itemize}
\item An adversarial entity can choose a specific individual and, with significant probability, retrieve some of their search records with high certainty;
\item An adversarial entity can choose a random search record and, with significant probability, reidentify the end user with high certainty;
\item An adversarial entity can perform attacks that reidentify a large number of end users with high certainty and with reasonably simple methods.''
\end{itemize}

Although the notion of residual likelihood of identification is mentioned several times in the Joint European Commission and EDPB draft guidelines on the DMA-GDPR interplay~\cite{interplay}, no definition of this notion is provided.
In fact, the above definition of ``beyond residual risk'' cannot be found in the EDPB's guidelines on anonymization~\cite{edpb} or the GDPR\cite{GDPR} or the DMA~\cite{DMA} or the interplay document~\cite{interplay}.  
In Recital (614) of~\cite{EC}, the EC acknowledges that this definition comes from a report on anonymization written by a single ``recognised expert in the field of anonymisation.'' However, neither the expert's name nor the report have been disclosed.

Beyond qualitatively defining residual risk 
with little justification, no precise quantification of it is given. Although
the EC's document explicitly mentions the notion of \emph{residual} or \emph{insignificant} risk more than a hundred times, the reader does not find how much such risk should be. Whether this accounts for a 1\% or 0.1\% (which, for the population of the EEA countries, corresponds to millions and hundreds of thousands, respectively) is left for interpretation.  
In any case, according to Article 6(11) of the DMA~\cite{DMA}, ``any query, click and view data [provided by the gatekeeper] that constitutes personal data shall be anonymised''. 
And according to Recital (61) of the DMA and Recital (26) of the GDPR, anonymous information refers ``to 
information which does not relate to an identified or identifiable natural person or to personal data rendered anonymous in
such a manner that the data subject is not or no longer 
identifiable.''
The latter formulation \emph{calls for preventing unequivocal reidentification in all cases}, rather than merely reducing the risk below some so-called and unspecified \emph{residual} threshold. In fact, all well-established privacy models proposed in the past 30 years have been specifically designed to prevent unequivocal reidentification of \emph{any data record}, without exception \cite{models}.

In this respect, in Recital (723) the EC acknowledges that ``while Alphabet’s testing has shown examples of reidentification, these have not been at scale,'' which means that the loose characterization of risk by the EC directly contradicts the anonymization requirement of Article 6(11) of the DMA.

In fact, data anonymization is especially about protecting the most vulnerable minority records --the records that are particularly exposed precisely because they are the rarest and most outlying \cite{muralidhar2025protecting}. This stands in direct contradiction with the EC's stated goal of preserving ``unique queries'' and queries in the ``long tail'' (Recital (572)).

\subsection{Ineffective anonymization}
\label{sec:noanon}

The suppression mechanism proposed by the EC in~\cite{EC} operates by removing queries that contain rare entities or words. However, it assesses each query component in isolation. Therefore, {\em it fails to consider whether the combination of all (rare or common) components makes the query itself unique and thus potentially reidentifying}. 

Regarding query metadata, although the EC requires at least 1,000 users to have issued queries sharing the same inferred language, generalized location, and device type, query timestamps are independently coarsened to a fixed daily granularity. Since a query's timestamp may act as a quasi-identifier, its independent generalization may undermine the protection afforded by the 1,000-user metadata coarsening.

More importantly, the query-suppression and metadata-generalization mechanisms operate independently of each other. Consequently, they provide no guarantee that unique search records --\emph{i.e.}, unique combinations of query content and associated metadata-- are eliminated or adequately masked. A record may therefore satisfy all the proposed thresholds for its individual components and metadata while remaining uniquely identifiable when these elements are considered jointly.

This approach is in conflict with the fundamental principles that underlie data anonymization (see Section \ref{sec:anonymization}). Unfortunately, this is not an isolated issue: the failure to account for the joint uniqueness of multiple attributes has been a recurring error in the data science literature, as discussed in \cite{science,sanchez2025anonymization}. However, in Recital (674) the EC justifies this design by stating that ``the query text does not need to be covered [...] to achieve $k$-anonymity. This is because [...] the query text should not be considered a quasi-identifier in the data sharing context of Article 6(11) of Regulation (EU) 2022/1925''. Also, in (678) the EC states that ``the exact query text issued by a given individual is unlikely to be an attribute about that individual that an attacker already knows, and can use in an attack''.

In fact, the notion of ``unlikely'' in (678) is rooted in the EC's inappropriate characterization of risk, which effectively assumes that anything short of straightforward and large-scale reidentification should not be regarded as a threat. 
Furthermore, this narrow view conflicts with nearly two decades of research on search query anonymization \cite{terrovitis08,He09,Navarro2012,Batet2013,Sanchez18}, which is grounded in the rather obvious observation that the query text may contain quasi-identifying information about the person who issued the query, even if such information occurs only infrequently.
For instance, 
queries submitted by the user about their neighborhood helped reidentify Thelma Arnold in the AOL database~\cite{aol}. 
As we show in the following, similar queries enabling reidentification, record linkage, and/or disclosure can still be retrieved after the EC's measures have been applied.
Thus, contrary to the position of the EC, the requirement set out in Article 6(11) —namely, that ``any query [...] that constitutes personal data shall be anonymised''— clearly applies in this case.

\subsubsection{Empirical analysis} 
To illustrate the arguments above, we counted the number of unique queries and unique search records remaining in the six-month Google Search dataset for Ireland after applying the EC's technical measures. The results are reported in Table \ref{tab:query-uniqueness}\footnote{Although some apparently unique queries in the Ireland six-month dataset might no longer be unique at the EEA level, their unequivocal association with the 
country of origin (or a more specific location in Ireland) removes any ambiguity.}.

\begin{table}[htbp]
\centering
\caption{Unique queries and search records in Ireland by time window.}
\label{tab:query-uniqueness}
\small
\resizebox{\textwidth}{!}{%
\begin{tabular}{@{}lr
                r@{\hspace{0.4em}}r
                r@{\hspace{0.4em}}r
                r@{\hspace{0.4em}}r@{}}
\toprule
 & & \multicolumn{2}{c}{\thead{Unique queries}}
   & \multicolumn{2}{c}{\thead{Unique search records\\without timestamp}}
   & \multicolumn{2}{c}{\thead{Unique search records\\with timestamp}} \\
\cmidrule(lr){3-4} \cmidrule(lr){5-6} \cmidrule(l){7-8}
\thead[l]{Time window} & \thead{Search records}
  & \thead{Count} & \thead{\%}
  & \thead{Count} & \thead{\%}
  & \thead{Count} & \thead{\%} \\
\midrule
1 day (2026/08/01)
        & \num{25724013}
        & \num{9037215} & 35.13
        & \num{11263041} & 43.78
        & \num{11263041} & 43.78 \\
1 week (2026/08/01-2026/08/07)
        & \num{195687166}
        & \num{53864967} & 27.53
        & \num{69994504} & 35.77
        & \num{87135099} & 44.53 \\
1 month (2026/08/01-2026/08/31)
        & \num{896671597}
        & \num{202464930} & 22.58
        & \num{267671390} & 29.85
        & \num{393681524} & 43.90 \\
3 months (2026/06/01-2026/08/31)
        & \num{2609775091}
        & \num{509856764} & 19.54
        & \num{683491602} & 26.19
        & \num{1138890630} & 43.64 \\
6 months (2026/03/01-2026/08/31)
        & \num{4871552575}
        & \num{872692810} & 17.91
        & \num{1182531997} & 24.27
        & \num{2134670108} & 43.82 \\
\bottomrule
\end{tabular}%
}
\end{table}

Note that we assessed the uniqueness of a query using the \emph{modified query} rather than the \emph{initial query}, as the former is less prone to typos and inconsistencies that could artificially inflate uniqueness rates. As expected, the proportion of unique queries decreases as the time window increases, because the search data become more
homogeneous. Nevertheless, even in the longest time window (6 months), nearly 18\% of queries are unique, and this proportion is already quite stable and unlikely to decrease significantly in the 13-month window defined by the EC to build the allow list. Uniqueness increases significantly when we consider search records (query + metadata). Taking into account only the three metadata attributes used by the EC for generalization (location, device type, and inferred language), the proportion of unique search records in the six-month window exceeds 24\%. Including the query timestamp, with independently fixed day-level granularity, raises this proportion to around 44\%, regardless of the time window considered. 

It is clear that \emph{a significant number of 
search records 
can be isolated}. Moreover, given the near-exhaustive coverage of Google Search data (meaning that uniqueness within this dataset is quite likely to imply uniqueness across the entire EEA population), any such record containing (quasi-)identifying information \emph{is likely to be reidentifiable}.
For example, as we show in Sections \ref{sec:identifiers} and \ref{sec:examples}, simple search patterns allowed us to retrieve hundreds of unique queries containing names that passed the entity-based threshold and, in some cases, birth years. Combined with the query location obtained from the metadata, this information enabled us to unequivocally reidentify several individuals.

In addition to identifying unique search records, we can exploit query metadata to \emph{link several search records belonging to the same individual} with very high probability. Such linkage poses a much more serious privacy threat than individual queries, as it allows much larger amounts of personal information to be aggregated. For example, examining unique queries issued in the least frequently represented languages in the Ireland dataset, we found that all queries of several minority languages were concentrated on a single day within a weekly time window. This isolation strongly suggests that these queries were issued by the same foreign person or, at most, by a small group of people traveling together. By combining the query date, language, likely whereabouts of the query issuer, and query content, we were able to compile a fairly detailed personal profile, as reported in Section \ref{sec:examples}.
Similar analyses can be easily automated for minority languages in each EEA country.

Search record linkage does not even require query metadata: the query content itself can enable unequivocal linkage when it contains specific identifying information. For instance, as we discuss in Section \ref{sec:identifiers}, many Irish Eircodes (consisting of 3+4 alphanumeric characters that identify postal addresses of individual properties) passed the implemented entity detector. As a result, we were able to retrieve thousands of queries containing Eircodes, many of which corresponded to private residences. Dozens of these Eircodes appeared in several queries asking for directions as the origin address, enabling us to \emph{link} the corresponding search records and aggregate personal information (\emph{e.g.}, interests and whereabouts) across several queries. As shown in Section \ref{sec:examples}, this enabled us to unequivocally reidentify individuals, link their search records, and compile personal information about them.

\subsection{Ineffective suppression of identifiers}
\label{sec:identifiers}

Having dismissed the need to mask quasi-identifiers within the query text, the EC argues that suppressing queries containing direct identifiers (\emph{e.g.}, person names, addresses, or telephone numbers) issued by less than 50 users is sufficient to prevent reidentification. 
As discussed above, this is far from the case: virtually any piece of information contained in a query may act as a (quasi-)identifier, and its frequency in isolation says little about its potential to enable reidentification when considered in combination with other information in the search record. Moreover, the fact that an identifier (\emph{e.g.}, a person's name) has been searched for by more than 50 individuals (\emph{e.g.}, friends or relatives) does not mean that it cannot still enable unique reidentification, since the identifier may refer unambiguously to a single individual.
In this respect, Recital (645) states that ``a name or address included in a query often [referring to the entity-based threshold of 50 occurrences] does not comprise end user's personal data'', because they are ``typically known to a larger circle of individuals''. The EC justifies this assertion not on the basis of prior research, empirical evidence, or its own assessment, but solely on the opinion of an undisclosed market participant: ``even if a query contained a name, there is no way to know who searched it''. This is an extremely unconvincing and naive argument: as we show in Section \ref{sec:examples} and in the empirical analysis below, queries containing a name can be identified as self-searches with a high degree of confidence when expressed in the first person (``my name...''). Similarly, an address included as the origin in a query requesting directions is very likely to correspond to the query issuer's residence, particularly when the same origin recurs across multiple queries. Combining these query patterns with additional metadata, such as the query location, further removes any remaining ambiguity. 

The approach adopted by the EC is therefore more reminiscent of weaker data protection practices such as de-identification, understood as the mere suppression of direct identifiers. 
Such weaker practices cannot, by themselves, render personal data non-personal within the terms defined by the GDPR \cite{sanchez2025anonymization}. Yet even under this weaker model, the EC's measures fall short of meeting the basic requirements of de-identification. A fundamental principle of de-identification is that \emph{all} identifiers should be suppressed. In structured databases, where information is organized into well-defined attributes, it is generally possible to identify all attributes that may be direct identifiers and subject them to suppression. Free-text search queries are fundamentally different: the universe of possible identifiers is open-ended and cannot be reduced to a predefined set of types.

In this respect, the EC relies on the ``use of personal data detectors to detect addresses, names, phone numbers, location coordinates, IBAN numbers, credit card numbers and other forms of well-known identifiers'' (Recital (597)). 
However, such detectors (commonly referred to as \emph{named entity recognizers} (NER)) are inherently imperfect,
that is, they never reach 100\% detection recall~\cite{nasar2021named,seow2025ner-review}.
Furthermore, the limitations of NERs are exacerbated in the context of search queries, where the syntax, spelling, formatting, and capitalization on which these detectors are based may be highly inconsistent~\cite{jeong2023consistency}. 
More fundamentally, {\em no predefined catalog of identifier types is guaranteed to capture the potentially unbounded range of information that may serve as an identifier in a completely open free-text query.}

Consequently, sooner rather than later, rare direct identifiers that do not conform to the expected patterns or identifiers belonging to categories not anticipated by the detectors will pass through the proposed safeguards. This is particularly problematic when each individual component satisfies the EC's entity-based threshold while the combination of components nevertheless constitutes a unique identifier, for example, a person's full name composed of several otherwise common words. 

Furthermore, as discussed above, not all direct identifiers are removed: only those that have been mentioned by less than 50 individuals during a period of 13 months are removed.

\subsubsection{Empirical analysis}
Alphabet's implementation of the EC measures uses Google Cloud Sensitive Data Protection\footnote{\url{https://docs.cloud.google.com/sensitive-data-protection/docs}} to detect 42 types of identifying and confidential entities through purpose-built detectors. These include all the entity types explicitly mentioned in the EC document, as well as many others (\emph{e.g.}, passports, MAC addresses, usernames, API keys, etc.). Although Google's Sensitive Data Protection is a mature, feature-rich enterprise service, with support for the large-scale data inspection required by this setting, its detection accuracy is inherently limited, as is that of any other NER-based tool. To illustrate these limitations and also the fundamental flaws of the deidentification approach adopted by the EC, we designed simple regular expressions that bypassed most of the identifier entity types listed by the EC in Recital (597):

\begin{itemize}
\item \emph{Person names}. Unlike artificial codes, person names are particularly challenging for NERs because they can easily be confused with 
common words (for example, many family names relate to trades), especially when they are not capitalized and appear in grammatically inconsistent contexts such as search queries. Person names are also highly ambiguous: a name may refer to many different individuals, and the fact that a ``popular'' full name passes the entity-based threshold does not imply that a particular individual bearing that name is especially famous or publicly exposed. A simple search for the pattern ``my name'' on the Irish dataset retrieved hundreds of queries containing full names that had escaped removal by the detector and the entity-based threshold (\emph{i.e.}, the complete names were not included in the allowlist, although their constituent words were), while minimizing ambiguity as to whether the names referred to the query issuers themselves.
Applying the same search to the Spanish dataset yielded several thousand queries containing highly specific personal names (first name + two surnames), which are highly identifying. By combining these detailed names with other information contained in the queries, such as birth year or location metadata, we were able to reidentify dozens of private individuals, as described in Section \ref{sec:examples}. 
\item \emph{Emails}. Email addresses are also mentioned by the EC in Recital (641) as potential identifiers and have been considered in Alphabet's implementation. 
Although well-formed ``rare'' email addresses were effectively suppressed, many non-well-formed unique ones escaped both the detector and the entity-based threshold. For instance, searching for unique queries containing the strings ``gmail'' and either `` @ '' or `` at '' returned thousands of Gmail addresses, hundreds of which consisted of a person's name split into ``name'' and ``surname'', followed by numbers (many times resembling a birth year), and ``at gmail.com''. The numbers attached to the person name probably confused the name detector, while individual names and surnames (or even common full names), evaluated in isolation, easily passed the entity-based threshold. Thus, not only were personal email addresses retrieved, but also, sometimes, the full name and birth year of the individuals to whom they belonged. This information, together with the location of the query, allowed us to reidentify several individuals, as detailed in Section \ref{sec:examples}. 
\item \emph{Addresses}. There are many ways to refer to addresses, either as standard postal addresses or as codes. For example, as introduced in Section \ref{sec:noanon}, Ireland has recently implemented the Eircode system. Although Google Cloud Sensitive Data Protection includes a specific detector for Eircodes, a simple regular expression matching sequences of three and four alphanumeric characters separated by arbitrary whitespace returned thousands of unique queries containing Eircodes, many of them below-threshold (\emph{i.e.}, not present in the allow list). 
The retrieval of below-threshold Eircodes is quite probably due to the fact that Eircodes can easily be mistaken by detectors for product codes, such as automotive or technology model and part numbers, while each group of three or four characters, considered in isolation, was sufficiently general to pass the entity-based threshold. Moreover, queries containing multiple Eircodes (\emph{e.g.}, requests for directions between locations) can confuse detectors if codes are not clearly separated. To minimize ambiguity on the query issuer, we specifically searched for that type of queries through the pattern Eircode\_expression + ``to'' + Eircode\_expression. When the origin Eircode identifies a private residence, the query is highly likely to have been issued by someone living there. As we further develop in Section \ref{sec:examples}, these queries are particularly problematic because they are both reidentifying, confidential (the destination address reveals the interests and whereabouts of the query issuer) and enable search record linkage if the same origin Eircode appears accross several queries. 
Similar fine-grained address codes composed of short alphanumeric strings that can easily pass the entity-level threshold are also used in other EU countries, such as the Netherlands.
\item \emph{Phone numbers}. NERs can detect well-formed (\emph{i.e.}, compacted) phone numbers with high accuracy. However, phone numbers are often arbitrarily split into groups of digits with variable length. A regular expression matching four groups of three digits separated by arbitrary whitespaces returned several private phones. In these cases, the whole number escaped the detector, and each group of three digits was general enough to pass the entity-based threshold. This enables attackers to compile private phone numbers and add them as contacts on their smartphone. If the phone number owner is registered with a messaging app such as WhatsApp, the attacker may be able to discover the owner's name or username, profile image, and other profile information, thereby achieving reidentification. 
\end{itemize}

Note that our analysis was conducted under conditions particularly favorable to entity detectors, as they generally achieve their highest accuracy in widely represented languages, for which more training data are available \cite{hu2020xtreme}. For less common languages, particularly those using non-Latin alphabets, the accuracy of language-specific detectors can only be lower.

One might argue that the entity detection failures reported above could be easily addressed by incorporating the regular expressions used in our analysis during the creation of the allow list. However, most of these expressions target malformed identifiers precisely because they do not conform to the patterns that existing detectors are designed to recognize. Since the range of possible malformed representations is effectively unbounded, attempting to capture them through additional patterns would inevitably introduce numerous false positives (as we certainly observed), leading to the suppression of legitimate queries and a substantial loss of utility. This would be difficult to reconcile with the EC's prioritization of utility at almost any cost. More fundamentally, no predefined catalog of identifier types can exhaustively cover all the information (past, present, or future) that may serve as an identifier in unrestricted free text. 

All in all, the EC's deidentification approach thus rests on \emph{two assumptions that are impossible to meet in practice}: that all relevant identifier types can be anticipated in advance and that all instances of these identifiers can be reliably detected. 
Moreover, it also assumes that ``popular'' identifiers (\emph{i.e.}, those that have passed the entity-based threshold) cannot be disambiguated. As we have discussed here and document in detail in Section \ref{sec:examples}, simple search patterns can achieve precisely that.

\subsection{Arbitrary length-based threshold}
\label{sec:length}

The CE informally argues that longer queries pose a greater risk of reidentification because they are more likely to contain personal data (Recital (662)). In more precise terms, longer queries have a greater chance to contain larger combinations of unique quasi-identifiers and potentially confidential information, thereby facilitating both identity and attribute inference; but this is not a certainty, as disclosure via quasi-identifiers ultimately depends on whether the query is unique, regardless of its length. To address this alleged risk, the EC introduces an arbitrarily selected length-based threshold that suppresses search records containing queries longer than 95\% of the distinct queries issued in 2025 in the inferred query language.

At the same time, in Recital (637g) they specify that the dataset must include the first query entered by end users in AI mode. AI mode queries are gaining popularity, and many users issue them from the search engine instead of explicitly accessing an AI chatbot. These queries are characterized by being long and rich in details on the end users' context because this context is what allows chatbots to provide more accurate results. 
The relative volume of this type of queries can only be expected to grow in the (very) near future --a recent study highlights that the vast majority of end users are satisfied by the AI responses generated by their queries to search engines \cite{Chapekis25}). Thus, as this type of query increases in volume, two situations can arise:
\begin{itemize}
\item If the length-based threshold is fixed \emph{once} based on the length of queries issued in 2025, as indicated in 
Recital (598), then an increasing proportion of long queries (unique or not)
will be suppressed, resulting in increasing and uncontrolled degradation of data utility;
\item If the threshold is updated periodically (as stated in the EC's preliminary proposal \cite{preliminary}), then it will tend to grow because AI mode queries are likely to grow both in quantity and in length over time; in turn, this would cause nearly all non-AI model queries to be shorter than the growing threshold, which would render this threshold-based protection ineffective, not only in filtering out longer traditional queries, but also AI-oriented queries/prompts, which are much more disclosive than traditional search queries \cite{Paz25}.  
\end{itemize}

Both issues stem from the arbitrary nature of the length-based threshold proposed by the EC: queries should not be suppressed or masked simply because they are long —a long query may still be shared by many users— but because they are potentially disclosive, that is, unique.

\subsubsection{Empirical analysis}
We have analyzed the lengths in characters of distinct English queries over six months in the Ireland dataset, from March to August 2026. The average monthly lengths are 28.41, 28.47, 28.56, 28.84, 28.95, and 28.96 characters. This shows a slight but steady increase in length over time. The respective standard deviations are 11.31, 11.34, 11.35, 11.41, 11.44, and 11.44, which also increase slightly. 
This is compatible with longer queries in AI mode becoming more frequent.

\begin{figure}
\includegraphics[width=\textwidth]{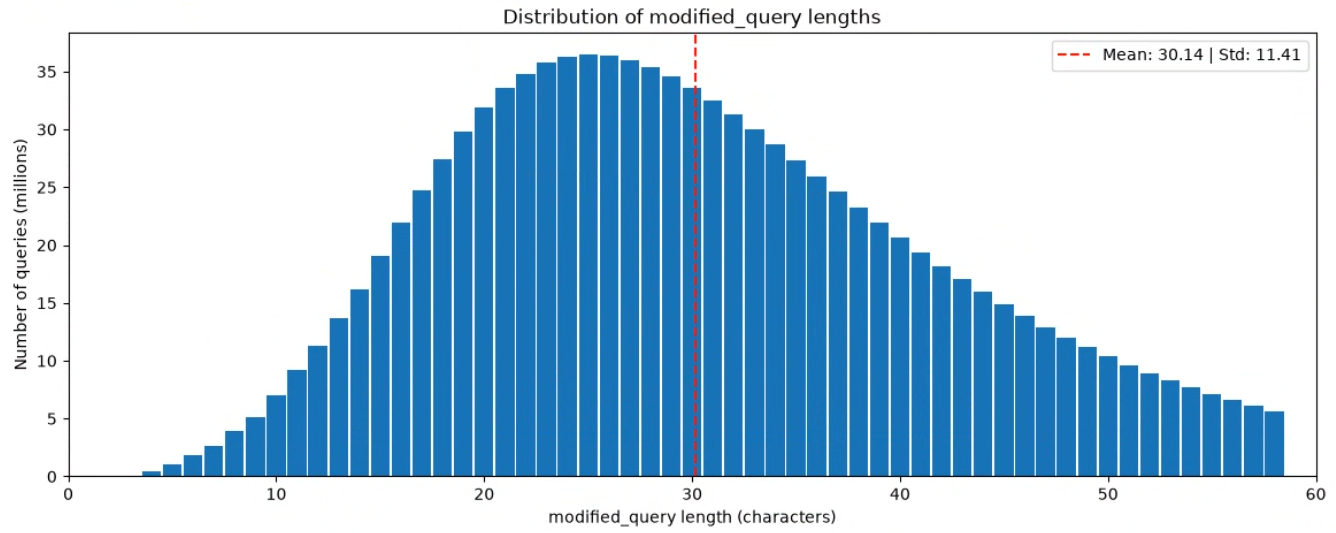}
\caption{Distribution of distinct English query lengths in the Ireland dataset from March to August 2026. Only queries with lengths at most 58 are retained.}\label{lengthireland}
\end{figure}

Figure~\ref{lengthireland} depicts the length distribution of the 
distinct queries within the six-month period. Its mean is $\mu = 30.14$, its standard deviation is $\sigma = 11.41$, and the right tail is truncated by the length-based threshold, which is 58 characters for English queries. The six-month mean is higher than the mean of every individual month because short queries tend to recur across months and are counted only once when the six-month datasets are pooled, whereas long queries are more likely to be specific to the month in which they were issued. Since the yearly data used to establish the length-based threshold is expected to yield an even higher mean, our analysis below should be considered conservative.

\emph{Lognormal estimation of the distribution of original query lengths.} The six-month distribution in Figure~\ref{lengthireland} and the one-month distributions have the shape of a lognormal distribution with a right tail truncated at a threshold $T$. We have also computed analogous histograms for other EEA countries during the same six-month period, and in all cases the histograms have a lognormal shape with a truncated right tail.
 
The mean and standard deviation of the lognormal distribution that fits the six-month distribution of the Ireland dataset {\em without truncating the right tail} can be estimated at $\mu = 33.66$ and $\sigma = 16.71$. This models the distribution of the length of the original distinct queries before discarding those longer than 58 characters. The goodness of fit of this distribution when truncated at 58 with the histogram in Figure~\ref{lengthireland} is $R^2=0.992$.

{\em Utility degradation for a fixed threshold as the average
query length increases}. If we fix a threshold $T$, the probability that a Lognormal$(\mu,\sigma^2)$ random variable $L$ representing 
a query length takes a value higher than $T$ is
\begin{equation}
\label{probexceed}
  \Pr(L > T) = \Phi\left(\frac{\mu' - \ln T}{\sigma'}\right),
\end{equation}
where $\Phi(\cdot)$ is the cumulative distribution function of the $N(0,1)$ distribution, and $\mu'$ and $\sigma'$ are the mean and standard deviation of the normal distribution underlying the lognormal, calculated as
\begin{equation}
\label{mumu}
\mu' = \ln(\mu) - \frac{\sigma^2}{2}, 
\end{equation}
\begin{equation}
\label{sigmasigma}
 \sigma'= \sqrt{\ln \left( 1 + \frac{\sigma^2}{\mu^2}\right)}.
 \end{equation}
 
Clearly, in Expression~\eqref{probexceed}, when the threshold $T$ is fixed (as in the final proposal of the EC), the probability of exceeding $T$ increases with the average length $\mu$, because an increase of $\mu$ results in an increase of $\mu'$, according to Expression~\eqref{mumu}. Hence, a higher proportion of queries will be suppressed as $\mu$ grows over time, leading to a diminishing utility of the protected dataset.
 
Now assume that $\mu$ remains constant and $\sigma$ increases. Since the length threshold $T$ is in general greater than the mean length $\mu$, from Equations \eqref{probexceed} and \eqref{mumu}, it follows that the argument of $\Phi$ in Equation \eqref{probexceed} is negative. When $\sigma$ increases, the denominator $\sigma'$ in the argument of $\Phi$ increases logarithmically and the argument becomes less negative; therefore, the value of $\Phi$ increases, and so does the probability \eqref{probexceed} of exceeding the threshold.
Thus, increasing $\sigma$ also leads to a decrease in the utility
of the protected dataset.

\emph{Estimating length-based query suppression.} We do not have access to the original 2025 queries that were used to derive the 5\% threshold $T = 58$. We therefore extrapolate the monthly trends for $\mu$ and $\sigma$ observed in 2026 back to 2025, which yields a lognormal distribution with mean 31.24 and standard deviation 15.00. Under this distribution, 5.63\% of queries are longer than 58 characters, which is close to the 5\% target and consistent with our conservative estimation. With the threshold fixed at 58, as in the EC's proposal, the share of suppressed distinct queries rises to 8.16\% in March--August 2026. 

To assess the future effect 
of a growing average length on the proportion of suppressed queries, let us make the conservative assumption that the standard deviation stays
unchanged at 16.71.
If the average length keeps growing at the observed rate, it will reach 36.76 characters in 2027 and 39.53 characters in 2028.
 In this case, the fixed threshold would suppress 10.23\% of distinct queries in 2027 and 12.54\% in 2028, compared to 5\% in 2025. These conservative figures show a clear and increasing loss of utility.

Let us now assess the effect of a growing standard deviation.
Indeed, we observe that the values of $\sigma$ also grew during the six months from March to August 2026. At the observed rate, we can extrapolate that $\sigma$ will reach 18.91 in 2027 and 20.87 in 2028, and the fixed threshold would suppress 11.84\% and 15.36\% of distinct queries, respectively, that is, up to more than three times the intended 5\%. Thus, the loss of utility for a fixed threshold increases with both the mean and the standard deviation of lengths, and both statistics 
have a growing trend in the data.

\subsection{Improper characterization of confidential information}
\label{sec:confidential}

To break anonymity and contravene the GDPR, it is sufficient for a search record to be unequivocally associated with an individual's identity, regardless of whether it also discloses confidential information. Nevertheless, we next argue that the remaining information in the queries can also reveal confidential information.

In this respect, the EC explicitly states that information referring to entities other than the query issuer should not be subject to protection (see, among others, the Recitals (574) and (623)). On this basis, it criticizes the application of $k$-anonymity to the entire query text. However, the EC measures are not capable of distinguishing whether a given query refers to the query issuer or to other entities: instead, they systematically suppress queries containing rare words or entities, regardless of who they refer to.

Admittedly, distinguishing with the (near-)perfect precision required for meaningful privacy protection which parts of a query refer to the query issuer and which refer to other unrelated entities is not feasible in practice. But the point is that requiring such a distinction is itself misguided. Even when some content of a query refers to other entities or individuals (\emph{e.g.}, books, celebrities, medical treatments, etc.), the mere fact that the end user searched for that information may reveal sensitive personal information --including their interests, political opinions, religious beliefs, or sexual orientation, all of which constitute ``sensitive personal data'' under the GDPR. In this respect, the EC's document appears to adopt an unduly narrow conception of secret personal information, focusing primarily on explicit and highly specific ``secrets'', such as passwords or bank account numbers, while overlooking the privacy implications of \emph{any} non publicly exposed personal information that can be inferred from a user's search behavior.
A user's personal query can also compromise the privacy of other individuals, particularly those close to them, such as children, partners, or other family members. However, under the EC's narrow approach, the privacy of these third parties would not need to be protected.

On the other hand, the EC argues that queries are unlikely to contain both identifying and confidential information because the length-based threshold leaves insufficient room for both (see Recitals (751), (752), (770), among others). The EC illustrates this argument with addresses, which it claims consume most of the permitted query length. But in Section \ref{sec:identifiers}, we have shown that specific addresses can be identified through Eircodes, which require as few as eight characters. 
As we illustrate in Section \ref{sec:examples}, many queries with Eircodes also contain additional confidential information, such as destinations or personal interests.

The EC also argues that attributing any confidential information to a particular subject is highly improvable because self-searches are rare and that ``an attacker that assumes a name in a query text related to a self-searcher would be incorrect more than 99\% of the time''. This reasoning is overly naive: we have shown in Section \ref{sec:identifiers} that straightforward search patterns can be used to systematically retrieve thousands of self-queries. Furthermore, in Section \ref{sec:examples}, we show that many of these queries also disclose sensitive personal information, and that highly effective search patterns can be designed to target specific categories of confidential information.

Finally, the EC also appears to overlook the risks of disclosure arising from the aggregation of search records (illustrated by the AOL case) based on the assumption that, under its proposed measures, search records cannot be linked (Recitals (633) and (757)). However, as shown in Section \ref{sec:noanon}, search records can indeed be linked with high confidence, and their aggregation can reveal highly specific personal information, ranging from preferences and occupations to more sensitive details, such as the detailed whereabouts of query issuers (see detailed examples in Section \ref{sec:examples}).

\subsection{Unnecessary utility loss and biased utility loss assessment}
\label{sec:utility}

The crux of the discussion in the EC's document~\cite{EC} is the alleged harm that sound anonymization techniques such as $k$-anonymity would cause to the utility of the search data. Most, if not all, of the EC's decisions appear to be driven by the goal of preserving data utility at all costs --particularly the ``long-tail'' of rarer queries even at the expense of the privacy of their query issuers. In this respect, Article 6(11) explicitly requires the preservation of data utility by providing that anonymization should be performed ``without substantially degrading the quality or usefulness of the data.''

Unfortunately, the EC's method does not conform to
the usual good practice in anonymization, which specifically seeks to optimize the trade-off between privacy protection and utility preservation. This good practice can be {\em privacy-first},
which means enforcing a certain privacy model/guarantee such as $k$-anonymity with the least utility loss that the privacy guarantee allows; or it can be {\em utility-first}, which means applying one or more utility-preserving statistical disclosure control methods ({\em e.g.}, generalization, microaggregation,
swapping, etc. \cite{sdc-book}) and iteratively tuning their parameters with the goal
of preserving the maximum utility that is compatible with an acceptable level
of empirically evaluated privacy protection. 

In contrast, the EC's method systematically suppresses queries based on arbitrary filters and thresholds that are neither justified in terms of a formal privacy model nor empirically tuned. As a consequence, no privacy guarantees are achieved and there is no 
principled basis for determining how much utility should be sacrificed to achieve a given level of privacy. 

Given the EC's concern about data utility, queries could have been masked in many utility-preserving ways rather than being completely suppressed, including (i) \emph{partial suppression} of the query terms that make a query unique, (ii) \emph{partial swapping} of query terms between different but semantically similar queries to break unequivocal inferences, or (iii) \emph{total or partial generalization} of several unique queries to make them indistinguishable.

In fact, many of the arguments put forward by the EC about data utility do not originate in its own assessment, but rather in the views of undisclosed third parties or market participants (see Recitals (572), (575), (580), among others). However, while the EC dismisses arguments advanced by Alphabet or third-party opinions supported or commissioned by Alphabet on the grounds of a potential conflict of interest, the same standard is not applied to the self-interested market participants mentioned in the above recitals. 

All in all, and somewhat paradoxically, the anonymization method proposed by the EC incurs an unjustified loss of utility without achieving meaningful privacy protection, and it relies on questionable and quite likely biased assessments from third parties.

\subsection{Disregard of low-effort but powerful privacy attacks}
\label{sec:attacks}

Privacy attacks may be used to reidentify specific records and disclose the sensitive information they contain. Throughout this paper (and in Section \ref{sec:examples} in more detail), we have proposed automatable search patterns that can be used to carry out such attacks. 

The EC acknowledges the possibility of such attacks but appears to downplay their effectiveness (see Recitals (624), (625), (655) and (739), among many others), either on the grounds that they would require substantial effort from an attacker, their lack of certainty, or because the additional information they could reveal is supposedly limited. 

To illustrate how naive this argumentation is, let us consider the following straightforward \emph{targeted attack}\footnote{According to Recitals (619) and (620), an adversary performs a \emph{targeted attack} when they can can ``choose a specific individual and, with significant probability, retrieve some of their search records with high certainty''. On the other hand, the adversary performs an \emph{untargeted attack} when they can ``choose a random search record and, with significant probability, reidentify the end user with high certainty'' or ``perform attacks that reidentify a large number of end users with high certainty and with reasonably simple methods''.}. Suppose an attacker possesses specific identifying information about one or more individuals --for example, their passport numbers, specific addresses, or email addresses-- which, due to their specificity, do not meet the threshold required for inclusion in the allow list. The attacker then deliberately issues queries containing this information from several simulated user accounts, artificially increasing its observed frequency. Since Google allows up to four user accounts to be associated with a single phone number, an attacker would need only 13 phone numbers (which can be obtained free of charge from many mobile operators) to generate the 50 account-level submissions required to manipulate the allow list.

Once the identifying information has been admitted to the allow list, subsequent legitimate queries containing that information would pass the entity-based threshold. The attacker could then filter out the queries generated from their own accounts and isolate the remaining legitimate ones. Since the targeted information is highly specific (otherwise, it would 
have appeared in the allow list prior to the attack), the attacker has strong grounds to attribute these remaining queries to the targeted individuals. In summary, the attacker has transformed rare information that was likely to be suppressed into an observable signal that can be used to identify the corresponding records and learn any additional information they might contain.

Moreover, according to the EC's proposed measures, entities admitted to the allow list remain there for five years. This substantially extends the attack window and increases the opportunity for an attacker to aggregate queries and confidential information over time.

Note that this attack would be ineffective if search data were protected using $k$-anonymity. For a query to meet the $k$-anonymity threshold, the attacker would have to inject $k-1$ instances of the \emph{exact same complete query}, together with metadata compatible with that of a query that the target user might subsequently issue. Even in the extremely improbable event that this was possible, the attacker would gain no additional information: the only records satisfying the $k$-anonymity requirement would be instances of the very same query that the attacker had injected in the first place.

In Recital (739), the EC acknowledges the possibility of this attack, but questions its effectiveness for user profiling with the following ingenuous argumentation: ``whether such attack is ``effortless'' [...] is debatable – an attacker would need to create fifty inauthentic accounts that avoid Alphabet’s filters for invalid traffic'' and ``it is not clear how such an attack would subsequently allow profiling''. 
We have argued that creating 50 accounts is not particularly difficult and can be done for free, and the effort involved can be amortized over an unlimited number of attacks/individuals and an unlimited time window. Moreover, the queries issued by the attacker can easily pass Alphabet's invalid traffic filters, which basically look for non-human automated searches submitted at high volume and speed, or improper interactions with the search interface (\emph{e.g.}, spam-like clicking behavior)\footnote{\url{https://support.google.com/google-ads/answer/11182074?hl=en}}. In this case, the attacker only needs to issue a few dozen realistically-looking queries for each target, either manually or through straightforward automation that mimics human search behavior. Regarding profiling, it becomes very plausible given the large amount of information that an attacker could accumulate over five years of continuous and probably unambiguous surveillance.

\subsection{Contractual measures cannot make data anonymous or compensate for inadequate anonymization}
\label{sec:contract}

The EC's technical measures are complemented by a set of contractual measures that essentially prohibit recipients from attempting to reidentify individuals. The EC explicitly presents these measures as a means to ``further mitigate that residual risk [arising from the technical measures] to an insignificant level'' (Recitals (618), (628), (648), (774), (830), (833) among many others). 

On the one hand, relying on contractual prohibitions in this way implicitly 
assumes that technical measures cannot, on their own, provide sufficient protection against reidentification, that is, \emph{they cannot effectively anonymize the data}.
On the other hand, this approach appears difficult to reconcile with Recital (26) of the GDPR, which provides that the assessment of privacy risks must take into account the means that are ``likely to be used'' for identification. The GDPR framework inherently contemplates the possibility that reidentification may be attempted, rather than assuming that attacks will not occur because they are contractually prohibited. In fact, no contractual obligation can provide effective protection against malicious insiders, accidental or deliberate data leakage, or other unauthorized forms of disclosure.

The EC's approach also clashes with the EDPB's position~\cite{edpb} that 
``technical, organisational and contractual measures can help to restrict access, but are not by themselves sufficient to make data anonymous; while
controllers are encouraged to adopt these measures wherever appropriate, they should still
examine the (re-)identification capabilities of anyone who might be able to access the data
with means reasonably likely to be used.''
The insufficiency of contractual measures is further justified in \cite{edpb}: ``contractual measures can typically be subject to revision, or may even be disregarded entirely by the
parties.''

\section{Attack patterns and examples of reidentification, confidential attribute inference, and query linkage}
\label{sec:examples}

In the previous sections, we have sketched several attack patterns that yielded reidentification, linkage, and/or attribute inference. In this section, we exemplify some of them in detail. All of the attack patterns we describe are reproducible and automatable, and result in numerous reidentifications with almost total certainty.

{\em The concrete examples of disclosure that follow 
should be taken as more serious leaks than the reidentification of Thelma Arnold that resulted the AOL dataset scandal}~\cite{aol}. The reason is that the AOL dataset was only
pseudonymized, which is known to be unsafe \cite{sanchez2025anonymization}, while the query
data we attacked had allegedly been ``anonymized'' according to the measures proposed by the EC.

\subsection{Attack pattern 1: Reidentification 
and confidential attribute inference through queries about oneself}

These are queries that contain ``I'' plus a verb, ``my'' followed by a noun, or other expressions indicating that the information sought concerns the query issuer (\emph{i.e.}, they are self-queries).
The entire content in such queries can be taken as attributes of the user making the query with high certainty, which helps reidentification and attribute inference. 
To facilitate reidentification, we specifically searched for queries including the query issuer's name with the pattern ``my name''.
A common feature of these queries is that users seem to expect answers about themselves from the AI-enabled search engine.

We conducted this attack against the six-month data from Spain, where most queries were in Spanish. Reidentification was facilitated by the use of two surnames by Spanish citizens, many of which escaped entity detection for the reasons explained in Section \ref{sec:identifiers}.

We applied and automated this \emph{untargeted attack} pattern as follows:
\begin{enumerate}
\item We searched for unique queries containing the Spanish equivalents of ``my name'': \emph{``me llamo''} and \emph{``mi nombre''}. This yielded 421,314 search records.
\item We submitted them to the Google Gemini API (gemini-2.5-flash), instructing the model to rank the queries according to the rarity of the full names they contained, and also priotizing any location-related information that might be mentioned. 
\item We manually examined the first 100 top ranked records, matching individuals' names and query metadata against publicly available online sources to confirm their identities. Thus, the examples we mention in the rest of this section are just a small fraction of what could be found.
\end{enumerate}

\subsubsection{Examples.} We were able to reidentify around a dozen individuals among the 100 records examined. In the following, we present some examples of these reidentifications that also resulted in the disclosure of confidential information. Original queries have been translated from Spanish into English, with identifying details generalized or redacted using []:
\begin{itemize}
\item QUERY: ``LENS\_IMAGE\_INPUT\_PLACEHOLDER Make me a similar diploma with my name [First-name Family-name1 Family-name2]''. Due to the fact that the two family names of this woman are relatively unusual, we found a unique match in the results of a public examination file for a 
position in [Spanish province capital]. The query location corresponded indeed to the capital's province. Not only did we reidentify her, but we also discovered the {\em confidential information that she wanted help in forging a diploma resembling the one provided as an image attachment to the search query}.
\item QUERY: ``My name is [First-name Family-name1 Family-name2] and I 
have suffered mistreatment''. This query was submitted from [Spanish province]. We have reidentified this woman as a registered lawyer in [Spanish province capital]. {\em The confidential information that she has suffered 
mistreatment is related to her sentimental private life and therefore qualifies
as special category data according to the GDPR.}
\item QUERY: ``My name is [First-name] and my boyfriend is [First-name2 Family-name1
Family-name2] where can I see him''. We were able to find the boyfriend in
[Spanish province] through a Facebook account. The query metadata confirmed
that it had been sent from [Spanish province]. In this case, we were able to reidentify the
boyfriend but not the woman who had submitted the query because she only gave
her first name. However, {\em this query leaks confidential information on the relationship
between that woman and the reidentified boyfriend: it suggests lack of trust, and such information might also qualify as special category data.} In line with what we mentioned in Section \ref{sec:confidential}, this is an example of a personal query that compromises the privacy of a third party. 
\end{itemize}

\subsubsection{Boosting this attack pattern.} This pattern has enabled us to achieve {\em reidentification at scale}: more than 10\% success in reidentification among the examined records. However, the effectiveness 
of this pattern and the likelihood of finding queries that also disclose confidential information can be further increased in several ways. 

For example, from the previous analysis, we discovered that some people make queries related to numerology and give their names and other information as input. {\em Note that the belief in pseudoscience such as numerology is a confidential attribute that is special category data according to the GDPR.} Thus, {\em any reidentification in this kind of queries also implies confidential attribute disclosure}. Moreover, such information is frequently exploited for scams. To do a more targeted search, we filtered the search records matching the ``my name'' pattern by looking for any of the following keywords: \emph{``numerologia''} (numerology), \emph{``cabala''} (Kabbalah), \emph{``numero''+``alma''} (soul number). This produced 594 results, which were ranked by Gemini according with the same prompt as above.
This pattern was highly successful: all the 15 highest-rank queries contained full names, without false positives, and we were able to unequivocally reidentify five of the individuals concerned (a 33\% reidentification rate). These cases are summarized below:
\begin{itemize}
\item QUERY: ``Determine the numerology and Kabbalah for my name [First-name Family-name1 Family-name2]''. Since [First-name] is a rather rare female name and [Family-name2] is a long name that is not that usual, we found a unique match who is a manager of a company 
headquartered in [Spanish province capital]. The location of the query indeed corresponded to the capital's province.
\item QUERY: ``My name is [First-name Family-name1 Family-name2] what is my soul number''. [Family-name1] is not a very common surname. We found a unique match in the public records of the government of the [Spanish province capital], 
working in the public health system of [Spanish province]. The location of the query was indeed [Spanish province].
\item QUERY: ``What is the soul number corresponding to my name [First-name Family-name1 Family-name2]''. We found a single match for the name: the LinkedIn profile of a woman who works in a telecommunications company 
in [Spanish province capital]. Furthermore, the location of the query shows that the query was sent from the capital's province.
\item QUERY: ``And what does Kabbalah suggest about my name [First-name Family-name1 Family-name2]''. 
We found only one individual with that name listed as a partner in a business operating in the hospitality sector in [Spanish province], precisely where the query was issued.
\item QUERY: ``According to numerology my name is [First-name Family-name1 Family-name2]''. The first name was compound, and the family names were relatively uncommon. We found only one matching individual, who owned a 
shop in [Spanish province capital], the location indicated by the query.
\end{itemize}

A broader category of confidential information is related to health, which is also classified as a special category of personal data under the GDPR. We targeted this type of information by instructing Gemini to prioritize queries containing both uncommon full names and health-related information, including mental health conditions, stigmatized diseases, cancer, sexual and reproductive health, sexual dysfunction and infertility, substance use, neurological disorders, genetic and autoimmune diseases, metabolic, gastrointestinal and urinary disorders, rare and disabling conditions, pregnancy loss, and gender reassignment. As in the previous analysis, we examined the first 
100 records and identified the following cases involving both reidentification and disclosure of confidential information:

\begin{itemize}
\item QUERY: ``My name is [First-name Family-name1 Family-name2] do I have anxiety?''. This query was submitted from [Spanish province capital], which matched the query location.
We reidentified this man from his LinkedIn profile; in the open part of it, he mentions that he works
in the National Defense in [Spanish province capital]. We also found his Instagram page with military-related activities.
{\em The fact that this man is concerned about his mental health is confidential information in the special categories of the GPDR. The disclosure is all the more serious for at least two aggravating circumstances: first, this man works 
in Defense, and second, he has access to guns.}
\item  QUERY: ``I need you to simulate a medical leave of absence under my name
[First-name Family-name1 Family-name2]''. According to the query metadata, this
query was sent from [Spanish province]. The name corresponds to a woman who was one of
the instructors in a master’s degree whose syllabus we found on the web. In that
syllabus, it was mentioned that the woman works in a company based in [Spanish province capital]. {\em Like
a previous example, this query suggests that the user wants help in forging a document, in this case a medical leave. This is even more sensitive than the previous query about diploma forgery, because forging a medical leave could entail penalties from the employer or even job loss.}
\item QUERY: ``My name is [First-name Family-name1 Family-name2] when I had prosthesis
surgery''. The family names are not very common, especially [Family-name2]. We
were able to find this person in official documents from [Town in Spanish province]. The location in the query metadata confirmed that the query
was sent from [Spanish province]. {\em This query contains health information on the user, which is special category data according to the GDPR.}
\item QUERY: ``Checking on a Prostalis order my name is [First-name Family-name1 Family-name2]''. 
This query was sent from [Spanish province]. We found that this man had run a marathon about 20 years ago
in [Town in the Spanish province]. 
We also found a traffic offense associated with him in another year, recorded 
near where he ran the marathon. 
Although traffic offenses may appear sensitive, they are public information. {\em What are really confidential data in the special categories of the GDPR is the order of Prostalis, which is advertised as a plant-based dietary supplement for prostate health and urinary comfort and has been surrounded by controversy over allegedly fraudulent marketing.}
\end{itemize}

Similar strategies can be designed to increase the success of the attack and/or target it toward a specific type of sensitive information (\emph{e.g.}, religion beliefs, sexual orientation, etc.).

Note that keyword-based filtering is straightforward to implement and inexpensive to run, and that Gemini processing costs for ranking queries as we have described are just  $\sim\$1$ for 10,000 queries. Consequently, all the experiments reported in this section, which exhaustively processed  country-level matching data, incurred a total cost of less than \$50.

\subsection{Attack pattern 2: Reidentification 
and confidential attribute inference through email addresses}

Many users send queries about problems with their email accounts. In some cases, they even naively ask Google to tell them their passwords. 

As introduced in Section \ref{sec:identifiers}, we searched for this type of query using the pattern ``gmail'' combined with either  ``@'' or ``at'', targeting malformed email addresses that may have evaded the email detector. This search returned 25,687 search records with unique queries in the Spanish dataset. Given the regularity of these patterns and common user habits, many of the retrieved email addresses included first names, surnames, and birth years, significantly facilitating reidentification. Then, we searched for queries that included the word ``password'' (``contraseña'' in Spanish) and analyzed the 50 longest ones. 

\subsubsection{Examples.} In the following, we report some reidentification examples, translating the original Spanish queries into English, and generalizing or redacting identifying details. All email addresses were validated at \url{https://email-checker.net/check}:

\begin{itemize}
\item QUERY: ``[First-name Family-name1 Family-name2 Two-digits]@gmail.com change password''. We found on LinkedIn a man with the same first name and surnames. The open part of his LinkedIn profile includes his picture and the fact that he works for a 
company based in [Spanish province capital]. The query location confirmed that the query was sent from [Spanish province]. {\em The confidential information disclosed by the query on the reidentified user is his personal non-publicly exposed email address and that he does not know how to change his password, suggesting poor literacy in the use of e-mail.}
\item QUERY: ``What is the password for my email account [First-name Family-name Four-digits]@gmail.com''. The first name and family name were not very common, and we found a single matching LinkedIn profile of a woman with that name born in [Four-digits]. The woman is from [Spanish city]. The location of the query confirmed that it was sent from the city's province. 
{\em The confidential information disclosed by the query on the reidentified user is her personal non-publicly exposed email address and that she lost her 
password and does not know how to reset it, which also
suggests poor literacy in the user of e-mail.}
\item QUERY: ``See the password for [First-name Family-name Four-digits]@gmail.com''. 
We found Instagram and Pinterest profiles of a man with this name born in [Four-digits]. 
This query could have been made by the user born in [Four-digits] whose name is [First-name Family-name] or by someone else trying to hack the account's password. Thus, we do not know whether the man lost his password or someone else tried to steal it; in any case, the query issuer is not very expert because search engines cannot in general show passwords unless these are publicly exposed. 
\end{itemize}

For the last two examples, which contained only a family name, we were able to find the corresponding [First-name Family-name] combination in the allow list, suggesting that these names were shared by multiple individuals. Nevertheless, by combining the full name with the birth year and the location of the query, we were able to unequivocally reidentify both individuals. This illustrates the ineffectiveness of the allow list.

Note that, in all these examples, potential attackers have a clear basis for low-effort targeted phishing attacks, having learned: (1) the person's full name; (2) their birth year; (3) their approximate location; (4) their personal email address (which was not publicly exposed previously); (5) the specific password-related issue they experienced with their email account; (6) the exact day on which the issue occurred, from the day-level query timestamp; and (7) the person's apparent lack of familiarity with information technologies. These are therefore clear examples of queries that are both reidentifying and disclosive of confidential information, with the potential to cause direct harm to the individuals concerned.

\subsubsection{Boosting this attack pattern.} This pattern can be further enhanced by instructing an LLM to prioritize email queries containing full names, as we did for the self-search queries. A preliminary test showed that this straightforward filtering step eliminated all false positives among the 50 highest-ranked queries: all retrieved email addresses contained full names.

\subsection{Attack pattern 3: Reidentification, linkage, and attribute inference through multiple directions queries with the same origin}

As introduced in Section \ref{sec:identifiers}, the general pattern underlying this attack is to identify Irish Eircodes that avoided detection or passed the entity-based threshold in queries requesting directions, where the origin corresponds to a private residence. Furthermore, when multiple queries share the same precise origin, that origin matches the query locations and the query timestamps span an extended period of time, they are likely to have been issued by someone living at that address. This makes it possible to establish not only the user's whereabouts but also their identity, particularly when a family business is registered at the same address.
 
We conducted this attack against the six-month data from Ireland. We applied and automated this \emph{untargeted attack} pattern as follows: 
\begin{enumerate}
\item We used a regular expression to retrieve queries containing Eircodes: either a permitted letter (ACDEFHKNPRTVWXY) followed by two digits, or the special prefix D6W, optionally followed by a whitespace character or hyphen, and then four characters consisting of digits or permitted letters. This returned 713,586 search records, although a significant proportion were false positives mainly corresponding to product codes.
\item We programmatically grouped queries containing the same Eircode, resulting in approximately 304,585 groups, most of which contained a single query.
\item We then programmatically ranked the groups by the number of different Eircodes contained in the queries of the group. This allowed us to prioritize Eircodes appearing in multiple queries asking for directions to other Eircodes. Doing so also filtered out many false positives. However, we excluded groups containing more than 12 queries, as highly recurrent Eircodes are more likely to correspond to public places than to private residences.
\item Finally, we manually examined the 50 largest groups, focusing on Eircodes that repeatedly appeared as origins in queries requesting directions and that corresponded to private residences when searched on Google Maps.
\end{enumerate}

\subsubsection{Examples.} When necessary, we generalize or redact the query content with []:
\begin{itemize}
\item An Eircode [XXX YYYY] indeed corresponded to a detached villa in [Irish town]. By analyzing all queries in its group requesting directions from [XXX YYYY] to other places, we found that four had been issued in March 2026 and one in August 2026,
The March queries specified destinations in three Irish towns and at one hotel, while the August query specified a destination in a business park. 
Moreover, all queries had the same location, which matched the villa's location.
The fact that the origin Eircode corresponds to a detached villa, together with the long period covered by these queries and the matching locations, provided a good indication that the query issuer lived at the origin address. We then used Google and found that this villa had been the registered address of a family company that provides sports-related services.
The company's name included a surname. We found that the company 
had two shareholders. We also found a chat between the company manager and some customers, and we learned the manager's name, whose surname coincided with the surname in the company's name. 
Finally, we found a 
picture of the manager and his wife after a sports event, along with the name of the two of them. They were probably the two shareholders of the family company.
\item Another Eircode [WWW ZZZZ] corresponded to a detached private residence registered as the premises of a family-run company specializing in construction and environmental services. 
We were able to identify the owner's first and last name as the manager of the company in an open online register. 
Within the corresponding Eircode group, we found seven queries requesting directions from the company's premises to nearby private residences on specific dates over a three-month period. All queries shared the same location metadata, which matched the location of the origin residence. Again, this provided a strong indication that the query issuer lives at the origin address. The specified destinations were likely to correspond to customer visits.
\end{itemize}

In the two previous examples, {\em the confidential information disclosed by the query texts consists of the query issuers' detailed whereabouts, including specific locations and dates}.

In addition to these unequivocal reidentifications, we identified other query groups that contain potentially revealing Eircodes. For example, a group comprised about a dozen queries that originated from private residences and ended at a hospital. A similarly sized group included several queries originating from private homes and having an orthopedic store as their destination. Another group comprised around a dozen queries originating at a delivery company's premises and ending at private residences.

\subsubsection{Boosting this attack pattern.}
Three remarks are worth noting to illustrate how this pattern based on directions queries can be exploited much more powerfully:
\begin{itemize}
    \item All of the above steps used only Internet searches and LLMs available for free. An attacker can increase the number of reidentifications if they resort to paid databases. For five euros, anyone can access Ireland's property registry\footnote{\url{https://www.landdirect.ie/}} and find who owns a property. Similarly, anyone can find who are the shareholders of a company through paid access to several company registries available in Ireland.
    \item Manual verification of Eircodes can be automated using LLMs and Google Maps APIs to identify those corresponding to private residences and then search publicly available sources for the owners' identities.
    \item Whereas we did not have any background knowledge on the reidentified people, an attacker with some background knowledge would be able to reidentify many more people by performing \emph{targeted attacks} on specific Eircodes.
    \end{itemize}

\subsection{Attack pattern 4: Linkage through a minority language}

As introduced in Section \ref{sec:noanon}, query metadata can enable the linkage of multiple search records belonging to the same user. The attack pattern we consider here consists of identifying queries in minority languages that are clearly concentrated on a particular date.

\subsubsection{Example.} We searched for unique queries issued in the least frequent language in the Ireland dataset, which was 
Icelandic, and grouped them by day. Using these two straightforward filters, we found that \emph{all} $\sim50$ unique queries issued in Icelandic in March 2026 were concentrated in one single day. This concentration clearly indicates that all these queries were issued by, quite probably, a single Icelandic person. 

The queries were fairly consistent in content and clustered around several clearly defined topics (details have been deliberately omitted or generalized):
\begin{itemize}
\item A group of three queries were related to the same Icelandic newspaper;
\item A group of four queries were about an Icelandic professional football manager;
\item A group of two queries related to a murder committed on an Icelandic coastguard;
\item A group of three queries were about healthcare facilities in Iceland (two of the queries sought information
about the staff and medical personnel of the same center); an additional query asked about a specific prescription medication;
\item A group of two queries were about real estate and the property registry;
\item A group of three queries sought financial information related to the same Icelandic bank;
\item One query asked for the school calendar of a school, while another asked for the timetable of a supermarket, both in Iceland.
\end{itemize}

Moreover, the fact that all queries were issued in Ireland on the same day suggests that the query issuer was most likely traveling and may have submitted them while transiting through Dublin Airport and waiting for the day's only flight to Keflavík. Although the queries did not allow for direct reidentification, combining the date, language, likely whereabouts of the query issuer, and the content of the aggregated queries enabled us to compile a fairly detailed profile of that person, containing both quasi-identifying and confidential information. Such profiles can be exploited to mount background-knowledge reidentification attacks.

Other minority languages in the Ireland dataset whose queries were concentrated on a single day or, at most, two consecutive days on a weekly time window were Serbo-Croatian, Bengali and Finnish. A similar linkage attack could be carried out on queries in these languages and for other minority languages in other countries.

\section{The EC's measures do not satisfy any of the EDPB's anonymization criteria}
\label{sec:guidelines}

As a synthesis of the discussion and the results reported in the previous sections, we next explain why the EC's technical measures fail to satisfy any of the three criteria for effective anonymization set out in the EDPB's anonymization guidelines:

\begin{itemize}
\item \emph{No Record Isolation}. The EC's measures clearly fail to satisfy this criterion and, as shown in Section~\ref{sec:noanon}, 
they fail by a substantial margin: 
the percentage of unique search records is around 44\%, and this percentage translates to huge absolute amounts.
Unique search records can potentially be singled out even without background knowledge, as we have shown in Sections \ref{sec:identifiers} and \ref{sec:examples}. 
Moreover, our reidentifications were not obtained ``by chance'' or through manual analysis of millions of queries: all of them resulted from straightforward and automatable \emph{untargeted} attack patterns. This means that a significant number of reidentifications can be achieved with low effort.
\item \emph{No Linkage}. This criterion is not satisfied either. As demonstrated in Sections \ref{sec:noanon} and \ref{sec:examples}, search records can be linked not only to publicly accessible online sources such as LinkedIn and other social media platforms, but also to other search records belonging to the same user. Such within-dataset linkage is possible when records contain rare combinations of quasi-identifiers in the query text or metadata (\emph{e.g.}, minority languages), or recurrent or multi-word identifiers that evade entity detectors (\emph{e.g.}, Eircodes). Through this \emph{untargeted attack}, we were able to link dozens of search records and build detailed user profiles and mobility histories.
Moreover, as discussed in Section \ref{sec:attacks}, an attacker can easily construct a low-effort and powerful \emph{targeted attack} that systematically identifies and aggregates, with high accuracy, queries belonging to a particular individual. 
\item \emph{No Inference}. This criterion is not satisfied either. 
As discussed in Section \ref{sec:confidential} and illustrated in Section \ref{sec:examples}, search queries may contain numerous types of information capable of revealing confidential attributes of their issuers, many of them classified as special sensitive categories by the GDPR. 
The EC disregards such cases on the grounds that length-capped queries rarely contain both reidentifying and disclosive information and/or that it is not possible to determine whether the query content refers to the query issuer. We have shown that queries that are both reidentifying and disclosive are far from exceptional and that specific search patterns, such as self-searches, leave little room for ambiguity on the query issuer's identity. 
\end{itemize}

\section{Conclusions}
\label{sec:conclusions}

We have shown, both theoretically and empirically, that reidentification and confidential attribute inference may occur in Google Search data protected under the EC's technical measures. 
Rather than dismissing these cases simply because they are expected to be ``residual'' under the EC's proposed risk model, we argue that the risk model itself is fundamentally inadequate for assessing the privacy risks of this dataset: given the near-exhaustive coverage of the EEA population, even a small percentage of successful attacks may result in consequential disclosures about thousand of individuals. 
Moreover, we have challenged the EC's characterization of the reidentification risk as ``residual'' by demonstrating a variety of low-effort, automatable attack patterns that yielded numerous reidentifications of EEA citizens even on very small data samples.

To put this into context, it must be remembered that just one publicly reported reidentification on pseudonymized data was 
enough to trigger the AOL search data scandal~\cite{aol}; in comparison, the reidentifications we report are more serious, first, because they were achieved at scale and, second, because they involved search data that had allegedly been anonymized.

The conclusion is that \emph{the EC's measures cannot be regarded as effective anonymization}: they fail to anonymize the data in accordance with the requirements set out in Recital (61) the DMA and 26 of the GDPR, that is, to \emph{ensure that personal information does not relate to an identified or identifiable natural person}, and fail to satisfy any of the three criteria defined in the EDPB's anonymization guidelines.

The measures are also fundamentally heuristic and arbitrary, lacking even the empirical grounding usual in ``old-school'' SDC methods. In adopting this approach, the EC overlooks more than 30 years of research in data anonymization. 
This has two fundamental consequences. First, the resulting level of privacy protection is neither formally guaranteed nor rigorously bounded. Second, the resulting loss of data utility is not calibrated against any explicit or quantifiable privacy requirement, 
which is likely to entail a utility loss greater than necessary for
the (poor) privacy protection achieved.

Practitioners and regulators should therefore be cautioned against treating the EC's technical measures as an example of sound anonymization practice. Furthermore, the limitations of those measures cannot be remedied through contractual safeguards. Contractual restrictions may govern how authorized recipients use the data, but they cannot replace effective technical anonymization, compensate for weaknesses in the underlying data protection, or provide protection against malicious insiders, data leakage, or other forms of unauthorized disclosure.

Nevertheless, stakeholders can be assured that there are solid and well-established anonymization methods that can achieve a meaningful balance between a prescribed level of privacy and the preservation of data utility \cite{science}.

\section*{Acknowledgments and disclaimer}

Alphabet provided us with access to the data and the technical resources needed to conduct our analysis, as well as covered the expenses incurred in carrying it out. 
Partial support to this work was received from  
the Government of Catalonia (ICREA Acad\`emia Prizes to J. Domingo-Ferrer
and David S\'anchez), Agencia Espa\~{n}ola de Investigación
(project PID2024-157271NB-I0 ``CLEARING-IT''), 
and the European Commission (project
HORIZON-101292277 ``SoBigDataIP'').

The views and opinions expressed in this paper are those of the authors and do not necessarily reflect or represent the views or positions of Alphabet.

\bibliographystyle{splncs04}
\bibliography{bibliography}
\end{document}